\documentclass[prl,twocolumn,showpacs]{revtex4-1}
\usepackage{amsfonts,amsmath,mathrsfs,epsfig,amsbsy,bm,verbatim}

\newcommand{\bra}[1]{\left\langle {#1} \right \vert}
\newcommand{\ket}[1]{\left\vert {#1} \right\rangle}
\newcommand{\expect}[1]{\langle {#1} \rangle}

\begin{document}

\title{A proposal to probe quantum non-locality of Majorana fermions in tunneling experiments}
\author{Jay D. Sau$^1$}
\author{Brian Swingle$^2$}
\author{Sumanta Tewari$^{3}$}

\affiliation{
$^1$Department of Physics, Condensed Matter theory center and the Joint Quantum Institute, University of Maryland, College Park, MD 20742\\
$^2$Department of Physics, Harvard University, Cambridge, MA 02138\\
$^3$Department of Physics and Astronomy, Clemson University, Clemson, SC
29634
}

\date{\today}

\begin{abstract}
Topological Majorana fermion (MF) quasiparticles have been recently suggested to exist in semiconductor quantum wires with proximity induced 
 superconductivity and a Zeeman field. Although the experimentally observed zero bias tunneling peak and a fractional ac-Josephson
effect can be taken as necessary signatures of MFs, neither of them constitutes a sufficient ``smoking gun"
experiment. Since one pair of Majorana fermions share a single conventional
fermionic degree of freedom, MFs are in a sense fractionalized excitations.
Based on this fractionalization we propose a tunneling
experiment that furnishes a nearly unique signature of end state MFs in semiconductor quantum wires.
 In particular, we show that a ``teleportation"-like experiment is not enough to distinguish MFs from pairs of MFs, which 
are equivalent to conventional
 zero energy states, but our
proposed tunneling experiment, in principle, can make this distinction.
 \end{abstract}

\maketitle

\paragraph{Introduction:}
Majorana fermions \cite{Majorana} (MF) are localized particle-like neutral zero energy states that occur at topological defects and boundaries in superconductors. 
  A MF creation operator is a hermitian second quantized operator $\gamma^{\dagger}=\gamma$ which anti-commutes with other fermion operators. The hermiticity of MF operators
implies that they can be construed as particles which are their own anti-particles \cite{Majorana,Wilczek,Franz,Read-Green}. The key issues at this time in the condensed matter context are two fold, first, we must predict and characterize materials supporting MFs and second, we must detect them experimentally.  In this paper we address the second issue of experimental detection by proposing a nearly sufficent experimental signature for MFs.

MFs have recently been proposed to exist
  in  the topologically superconducting (TS)  phase of a spin-orbit (SO) coupled cold atomic gases \cite{zhang_c},
 semiconductor 2D thin film
 \cite{Sau,Long-PRB} or 1D nanowire \cite{Long-PRB,Roman,Oreg} with proximity induced $s$-wave superconductivity and Zeeman splitting from a sufficiently large magnetic field.
In principle, the MFs in such systems may be detected either by measuring the zero-bias conductance peak (ZBCP) from tunneling electrons into the end
MFs ~\cite{tanaka,Long-PRB,Sengupta-2001,R1,flensberg}, by detecting the predicted fractional ac Josephson effect \cite{Kitaev-1D,Roman,Oreg,Kwon,Fu-Frac,platero, meyer, nazarov,aguado}.
The semiconductor Majorana wire structure, which will be the system of our focus, is of particular present interest
 since there is experimental evidence for both the ZBCP  \cite{Mourik,Deng,Rokhinson,Weizman} and the fractional ac Josephson effect in the form
of doubled Shapiro steps \cite{Rokhinson}.

Despite their conceptual simplicity, neither the ZBCP nor the fractional ac-Josephson effect experiments constitute a sufficient proof of MFs at
the ends of topological superconducting wires.  A non-quantized ($2e^2/h$) zero bias peak, such as that observed in the recent experiments
 \cite{Mourik,Deng,Weizman}, can in principle
arise even without end state MFs \cite{Liu,Kells,Beenakker-Weak}. Similarly, 
 a fractional ac-Josephson effect can exist even in
Josephson junctions made of ordinary quasi-1D $p$-wave superconductors such as organic superconductors \cite{Kwon} or the non-topological phase of 
the semiconductor nanowire \cite{sau_bert}.
 Given these caveats as well as the considerable complexity of existing experiments, there have been several alternative proposals
 to detect the presence of MFs \cite{gil,demler,zoller,beri}. 
 Based on the inherent quantum non-locality of MFs, in this work we propose an alternative tunneling experiment on semiconductor Majorana wires that
furnishes a nearly sufficient signature of end-state MFs. We discuss in detail why only topological systems would show such 
quantum non-locality, which would even be absent for systems with conventional Andreev states at each end.

\begin{figure}
\centering
\includegraphics[scale=0.35,angle=0]{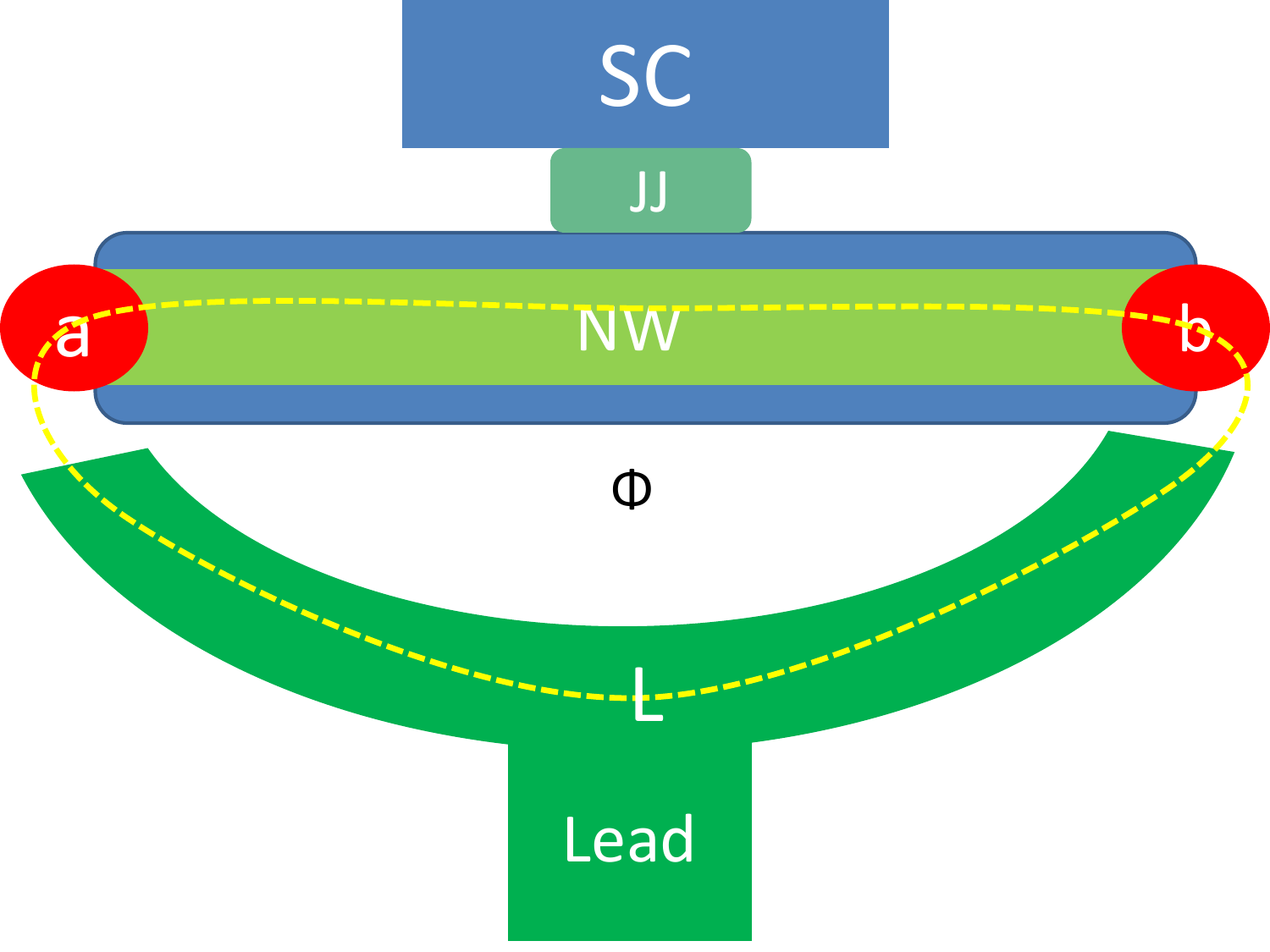}
\caption{(Color online)
Schematic experiment to detect Majorana-assisted electron tunneling between the MFs $a$ and $b$
along the dashed lines. The electrons in the ring are transported
via the Majorana-assisted tunneling process along the dashed line, which enclosed the flux $\Phi$.
The dark green metallic loop $L$ is connected to a metallic lead, which together with the superconductor serves 
as the terminals for the conductance measurement.
Similar to the Aharonov-Bohm oscillations in a mesoscopic ring, the interference can be detected as a $2\Phi_0=hc/e$ periodicity of the
 conductance  as a function of $\Phi$. The nanowire $NW$ is placed on a supercondcuting island (shown in light blue)
which has a charging energy $E_C$. 
}\label{Fig2}
\end{figure}

\paragraph{Non-local electron transfer}
Non-locality arises in MFs because they differ from conventional complex (Dirac) fermions in that they have no occupation number associated with them.
To define a quantum state of a system with MFs we must consider a pair of MFs.  The pair of MFs $\gamma_a$
and $\gamma_b$ at the ends $a$ and $b$ of a nanowire (NW) shown in Fig.~\ref{Fig2} can be combined into a zero-energy complex fermion operator
$d^\dagger=\frac{1}{2}[\gamma_a+i\gamma_b]$
associated with the pair of MFs $\gamma_a$ and $\gamma_b$ \cite{Kitaev-1D}.
The quantum state of the system is then determined by the eigenvalue of $n_d=d^\dagger d= 0,1$.  Since 
the fermion parity $F_P=(-1)^{n_d}$ associated with the operator $d^\dagger$ is related to the MFs by
\begin{equation}
F_P=(-1)^{n_d}=i\gamma_a\gamma_b,
\end{equation}
we see that the fermion parity
 of the whole system is determined by non-local correlations between the fractionalized MFs $\gamma_a$ and $\gamma_b$.
In fact, the fractionalization of the $F_P$ into a pair of spatially separated operators $\gamma_{a,b}$
 in one-dimensional systems with localized fermion excitations, is a unique characterization of the topological 
state of the system~\cite{berg}. 
  Our central concern is how to probe this non-locality to provide a robust and sufficient criterion of MFs.

An immediate idea involves trying to inject an electron into $\gamma_a$ and retrieve it from $\gamma_b$ ~\cite{sumanta,bolech,semenoff,fu}.
By connecting  leads to the left and the right ends of the TS wire in Fig.~\ref{Fig2}, one could imagine that an electron injected
into the end $a$ flips the occupation number $n_d$ from $n_d=0$ to $n_d=1$. The injected electron can then escape from the end
 $b$
flipping the state back from $n_d=1$ to $n_d=0$. Such a process where an electron can enter from one end $a$ and exit at the other
lead $b$, can be interpreted as a transfer of an electron, which we will refer to
as Majorana-assisted electron tunneling.
 However, as has been discussed in previous works \cite{sumanta,bolech}, such a transfer occurs in a way so as to not violate
locality and causality.

The amplitude for the Majorana-assisted electron tunneling \cite{sumanta,semenoff} can be written in terms of the retarded Green function as
 \begin{equation}
G^{R}_{mn}(\tau)=-i\expect{[c^\dagger_m(\tau),c_n(0)]}\Theta(\tau),\label{eq:A}
\end{equation}
 where $\tau$ is the time-interval between the tunneling events, $\Theta(\tau)$ 
is the Heaviside step function and
 $c^\dagger_{m}$ are the electron operators at the left (i.e. $m=a$) and right (i.e. $m=b$) end of the wire.
In the low-energy limit in the topological state, the electron-operators at the ends $c^\dagger_{a,b}$ can be approximated 
by the end Majorana modes $c^\dagger_{a,b}\sim\gamma_{a,b}$. Thus, the amplitude $G^R_{ab}(\tau)$ represents Majorana-assisted non-local  normal electron tunneling between the ends $a$ exits as a hole at the end $b$ 
and has a non-zero value in the topological phase given by
\begin{equation}
G^{R}_{ab}(\tau)=-i F_P,\label{eq:GR}
\end{equation}
 where  $F_P\equiv i\expect{\gamma_a(0)\gamma_b(0)}$ is the fermion parity of the 
 TS system. 
Since $G^R_{ab}(\tau)$ is directly related to the fermion parity $F_P$, 
the detection of such a non-vanishing amplitude for a non-local Green function $G^R_{ab}(\tau)$
 is a signature of the fractionalization associated with MFs. 

\paragraph{Charging energy:}
In equilibrium, the degeneracy of the different fermion parity states characteristic of a TS system lead to fluctuations in $F_P$, 
that would result in a vanishing average for the tunneling amplitude $G^R(\tau;ab)$. This is remedied \cite{fu} by introducing a
 charging energy $E_C$ on the superconducting island supporting the $NW$, which makes one of the fermion parities energetically 
favorable over the other.
To compute $G^R$, we consider the Hamiltonian for the NW in Fig.~1 as 
\begin{align}
&H=H_{BCS}+4 E_C (\hat{N}+\hat{n}_W/2-n_g)^2-E_J\cos{(\phi)},\label{eq:H}
\end{align} 
where $E_C$ is the charging energy of the wire, $H_{BCS}$ is the BCS Hamiltonian 
for the proximitized NW, $\hat{n}_W$ is the number of electrons in the NW.
 Here $\hat{N}$ 
is the total number of Cooper pairs with the SC island, the $NW$ and the gate and is a variable
 that is conjugate to the phase $\phi$, $n_g$ is the gate charge.
 To control the charging energy of the NW we couple it to a superconductor with
 Josephson strength $E_J$, which can in principle be controlled using a SQUID geometry\cite{SQUID}.

While coupling to the superconducting lead in Fig.~\ref{Fig2} breaks charge conservation, it preserves fermion parity 
$F_P$, which is related to the number of electrons modulo two \cite{vanheck}. 
 In the limit that $E_J\gg E_C$, so 
that the only effect of charging energy is an energy splitting $\delta=16\left(E_C E_{J}^3/2\pi^2\right)^{1/4}e^{-\sqrt{8E_{J}/E_C}}\cos{2\pi n_g}$
between the different fermion parity states $F_P=\pm 1$ . Thus the effective Hamiltonian is written as 
$H_{eff}=H_{BCS}+F_P\delta $. Since $F_P$ and $H_{BCS}$ commute, expanding $H$ in terms of $H_{BCS}$
the Green function $G^R(\tau)$ can be written as 
\begin{align}
&G^{R}_{mn}(\tau)=-i\Theta(\tau)\frac{\expect{[c^\dagger_m(\tau),c_n(0)] e^{-(\beta+i\tau) F_{P}\delta}}_0}{\expect{ e^{-\beta F_P\delta}}_0},
\end{align}  
where $\expect{\dots}_0$ is the thermal expectation with respect to $H_{BCS}$.

\paragraph{Coincidence probability:}
The amplitude $G^{R}_{ab}(\tau)$ can lead to a so-called coincidence
probability $P_c(\tau)$, which maybe measured by using a joint measurement by two point contact detectors at the two ends \cite{gong,sumanta}.
Alternatively, the non-local transfer of electron  can also be measured 
by a non-local conductance or transconductance  $\frac{dI_b}{d V_a}$ 
 between the ends $a$ and $b$ in Fig.~\ref{Fig2}. This measurement does not require closing the loop (L) in Fig.~\ref{Fig2} and 
would require adding a lead to the end $a$. 
In such a set-up, a voltage $V_a$ applied to the left-lead $a$ (relative to the SC) 
results in a current $I_b$ in the right lead $b$.
 Using results of Ref.~\onlinecite{bolech} for symmetric $t_a=t_b=t$ we find that
\begin{equation}
\frac{dI_b}{d V_a}=\delta \frac{32 V_a}{16 \Gamma^2+(\delta^2-V_a^2)^2+8\Gamma^2(\delta^2+V_a^2)},
\label{Paraconductance}
\end{equation}
which clearly vanishes for $\delta\rightarrow 0$ ( i.e. vanishing charging energy $E_C\rightarrow 0$).
 Here $\Gamma\propto t^2$ is the lead-induced broadening
of the MFs.

\paragraph{Topological versus non-topological systems:}
However, a coincidence measurement does not directly imply a non-zero $G^R(\tau;ab)$ in more general situations. 
 The amplitude $G^R_{ab}(\tau)$ in Eq.~\ref{eq:A},
 reflects the amplitude for being able to transfer an electron from $a$ to $b$ while leaving the state $\ket{g}$ invariant.
 On the other hand, the measurement of the
coincidence probability, $P_c$, does not keep track of the internal state of the system.
For a general system (i.e. one that may be topological or non-topological), $P_c$ 
for an electron entering at  $a$ and exiting at $b$ can be written more generally as
\begin{equation}
P_{c}(\tau)=\sum_{g_1,g_2}|\expect{g_2|c_b(0) c_a^\dagger(\tau) |g_1}|^2,\label{eq:P}
\end{equation}
where $g_1,g_2$ are the internal states of the wire, which are not necessarily identical.
While TS systems with MFs have a non-degenerate ground state in a given fermion parity sector,
more general systems with zero-energy Dirac end states may have multiple allowed values for $g_1,g_2$.
Therefore, the coincidence probability $P_c$ cannot be considered a unique signature for
a topological system.

An important example of the inequivalence of $P_c(\tau)$ and $G^R(\tau;ab)$ is a
non-topological superconductor with Andreev zero mode at each end.
  The quantum state is characterized by the occupancy
 $n_a,n_b$ of the two conventional zero energy end  modes.  We can easily have $P_c \neq 0$ in this non-topological setup.
Suppose the initial state is $g_1\equiv(n_a=n_b=0)$, then the sum for $P_{c}$ in Eq.~\ref{eq:P} would have a non-zero
 contribution from $g_2\equiv (n_a=n_b=1)$. The tunneling of an electron from the lead into the zero-mode at $a$ changes the 
occupation from $n_a=0$ to $n_a=1$. On the other hand, the electron required to change the occupation of the state $b$ from
 $n_b=0$ to $n_b=1$ comes from breaking of a Cooper pair. The other electron from the broken Cooper pair is 
emitted into the lead in the vicinity near $b$.
Note that the process conserves the number of electrons within the system and cannot be eliminated even by the
introduction of a finite charging energy \cite{fu}.
 Therefore in order to clearly distinguish this case from the
 process of Majorana-assisted electron tunneling (that also returns a non-zero $P_c$), we require $G^R(\tau;ab)$ given 
in Eq.~\ref{eq:A} itself to be non-zero.
  In other words,
 we require that the system return to the same state $g$ after the tunneling process, so the same electron that enters at $a$ leaves at $b$.
 The Green function between the ends of a non-topological systems, $G^R_{ab}(\tau)$, vanishes.
This is because introducing a superconducting phase-slip through a non-topological system which transforms $c_b\rightarrow -c_b$
and flips the sign of the Green function without affecting the Hamiltonian. In fact, in the Supplementary material~\cite{footnote} we explicitly show how this vanishes even in the case of decoupled pairs of MFs.

\paragraph{Proposed set-up:}
The Majorana-assisted electron transfer $G^R_{ab}(\tau)$ can be  measured by the setup in Fig.~\ref{Fig2} 
 consisting of an external semiconducting loop  (L) that is connected to the ends $a$ and $b$ of the NW.
The  Green function $G^R(\tau)$ can be determined  by 
measuring the Andreev conductance from the semiconducting loop $L$ into the superconductor shown in Fig.~\ref{Fig2} in the tunneling regime (i.e. small tunneling) 
with tunneling amplitude $t_{a,b}$ between the ends of the NW and loop. The tunneling Hamiltonian between L and NW is written as 
\begin{align}
&H_t= [t_a c_a^\dagger c_{L,a}+ t_b c_b^\dagger c_{L,b}]+h.c., 
\end{align}
where $c_{L,a},c_{L,b}$ are fermion annihilation operators in the loop $L$ near the ends $a,b$ and the 
flux affects $t_b$ as $t_b=t_{b,0} e^{i\varphi/2}$ where $\varphi=\frac{2\pi \Phi}{\Phi_0} $. 
 
The zero-bias conductance can be calculated using the Meir-Wingreen formula and expanding to 
lowest non-vanishing order in the tunneling amplitude as 
\begin{align}
&\sigma(\varphi)=\int d\epsilon \textrm{ sech}^2{\frac{\epsilon}{2T}}\sum_{m,n} Im[\Gamma_{m,n}(\epsilon)G^R_{n,m}(\epsilon)],
\end{align}
where $\Gamma_{m,n}(\epsilon)=t_{m}\rho_{mn}(\epsilon)t_{n}^*$ is the imaginary part of the lead-induced self-energy and the retarded 
Green function in the time-domain is written as $G^R_{n,m}(t)=\Theta(t)\expect{\{c_m(t),c^\dagger_n(0)\}}$.
Here the indices $m,n$ are summed over the ends $a,b$ and  $\rho=Im(G_L(0))$ is the density of states, which can 
be calculated as the the imaginary part of the retarded Green function in the loop.

Ignoring the energy dependence of the lead density of states $\rho_{mn}(\epsilon)$ and choosing (for simplicity) a symmetric lead  
and contacts with $|t_a|=|t_b|$, the imaginary part of the lead self-energy $\Gamma$ can be written as 
 $\Gamma_{aa}=\Gamma_{bb}=\Gamma_0$ and  $\Gamma_{ab}=\lambda\Gamma_0 e^{i \varphi/2}$ for appropriately chosen  
constants $\Gamma_0$ and $\lambda$. Within this set of simplifying approximations,  the conductance is found  to be  
\begin{align}
&\sigma(\varphi)=\int d\epsilon \textrm{ sech}^2{\frac{\epsilon}{2T}}\Gamma_0\nonumber\\
&Im\left[G^R_{a,a}(\epsilon)+G^R_{b,b}(\epsilon)+\lambda( G^R_{a,b}(\epsilon)e^{i\varphi/2}+e^{-i\varphi/2}G^R_{b,a}(\epsilon))\right].\label{eq:sigma}
\end{align} 
It is clear from the above formula that $\sigma(\varphi)$ shows $4\pi$-periodic oscillations whenever the 
non-local tunneling amplitude, $G^{R}_{ab}\neq 0$  across the $NW$ is finite. 
This direct measurement of the non-vanishing tunneling amplitude $G^{R}_{ab}$, which is a measure of the
 Majorana-assisted non-local tunneling process, would be a direct measurement of the non-local character of Majorana modes.
 
\begin{figure}
\centering
\includegraphics[scale=0.32,angle=0]{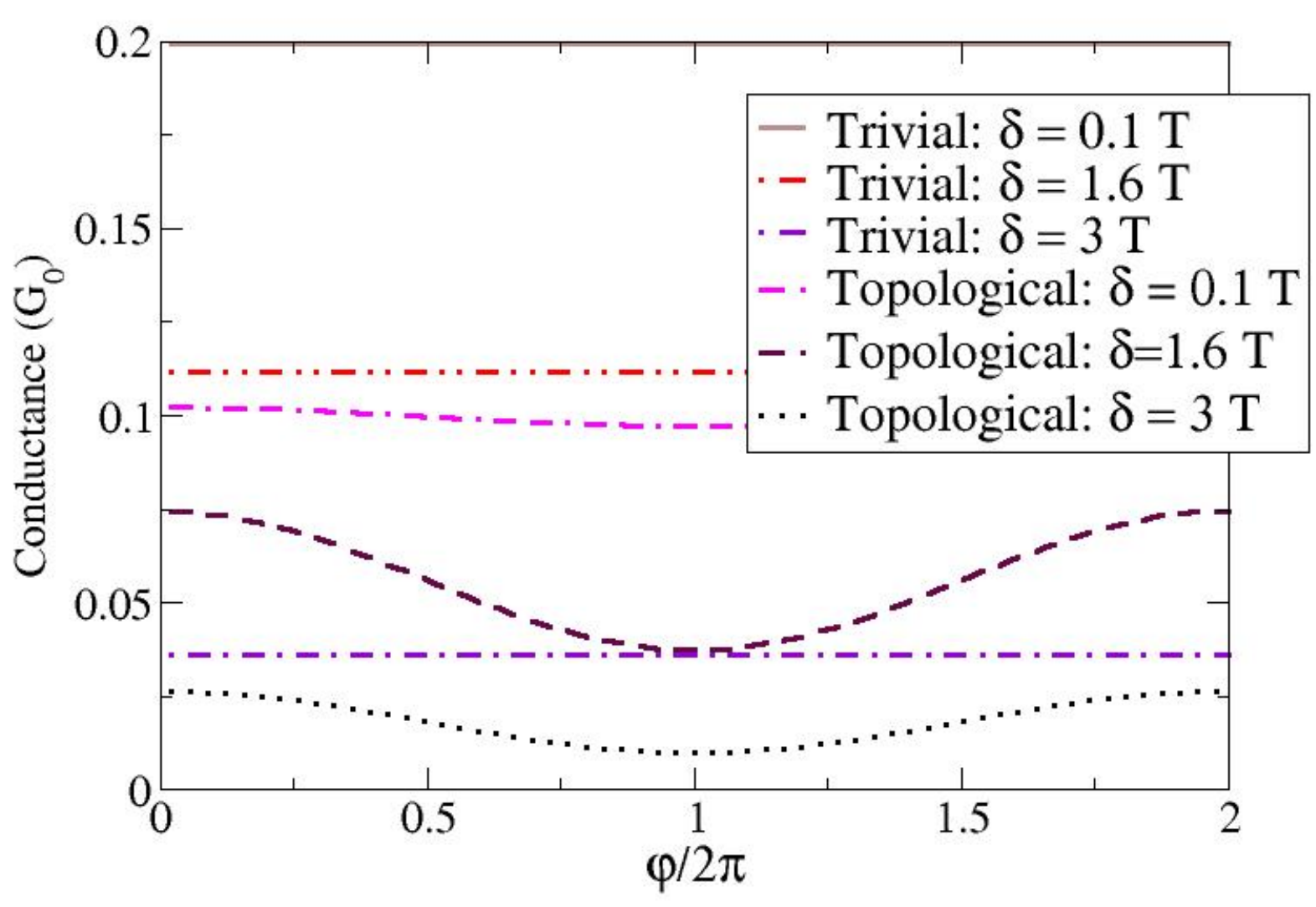}
\caption{
Tunneling conductance $\sigma(\varphi)$ as a function of flux $\Phi=\varphi\Phi_0/2\pi$ in the loop in Fig.~\ref{Fig2} 
in both the topological phase and the nontopological phase. Here $\Phi_0=hc/2e$ is the flux quantum and $G_0=e^2/h$ 
is the conductance quantum. The conductance shows $4\pi$-periodic oscillation in $\varphi$  
in the topological case that results from non-local transfer of electrons between the ends.
 In contrast, the spectrum in the nontopological case shows no dependence on phase at the lowest order in tunneling. 
}\label{Fig3}
\end{figure}

\paragraph{Results:}
To illustrate the periodic oscillations of $\sigma(\varphi)$ generated by the presence of non-degenerate Majorana modes,
  we calculate the conductance of  an 
InSb nanowire \cite{Roman,Oreg} with effective mass $m^*=0.015 m_e$, Rashba spin-orbit coupling
 $\alpha=0.5\,eV-\AA$, pair potential 
in the NW $\Delta=3\,$K. For simplicity, the loop is taken to be a semiconductor with effective 
mass $m^*$, but without spin-orbit coupling or Zeeman so that the spin-dependence of $\Gamma$ can be ignored.
 The chemical potential in the $NW$ is taken to be $\mu_{NW}=2\,$ K, while the loop is at a chemical 
potential $\mu_{loop}=6\,$ K. Further details of the model are provided in the Supplementary material~\cite{Supplementary2}.
The magnitude of the tunneling matrix elements $|t_{a,b}|$ are chosen to produce 
an experimentally reasonable zero-bias conductance of order $\sigma_0\sim 0.1 G_0$ at a temperature $T=50$mK where $G_0$
is the conductance quantum.
The conductance $\sigma(\varphi)$ in the set-up shown in Fig.~\ref{Fig2}  is plotted as a function of $\varphi$ in Fig.~\ref{Fig3} 
for both the cases where $NW$ is in the topological and non-topological phase. Details of the numerical evaluation of Eq.~\ref{eq:sigma} 
are provided in the Supplementary material~\cite{Supplementary3}.
The result in Fig.~\ref{Fig3}, shows that the conductance 
$\sigma(\varphi)$  including the charging energy shows a $4\pi$-periodic oscillation only in the topological case, as expected from 
non-local Majorana-induced quasiparticle transfer across the wire. 

The set-up in Fig.~\ref{Fig2} can be also  used to separate out 
Majorana-assisted electron transfer from direct transfer by tunneling of quasiparticles through $NW$.
In the topological case, the presence of a finite tunneling amplitude $G^{(R)}_{ab}$ depends on the charging energy
parameter  $\delta$, which can be controlled by a SQUID configuration \cite{SQUID}.
As seen in Fig.~\ref{Fig3}, the $4\pi$-periodic oscillations that are characteristic of the TS phase are completely 
suppressed for small $\delta/T$. In contrast, oscillations generated by direct tunneling of quasiparticles between the ends 
of the wire is not expected to be affected by $\delta$.

\paragraph{Comparison with the fractional Josephson effect:}
The signature of a TS phase in Fig.~\ref{Fig3} appears as a $4\pi$-periodic oscillations in conductance $\sigma$. 
Formally, this resembles the $4\pi$-periodic current-phase relationship predicted 
for  the fractional Josephson effect in TS systems\cite{Kitaev-1D,Sengupta-2001}. However, the $4\pi$-periodicity of 
the current in the Josephson junction in a TS 
system relies on fermion parity protection, which is typically accomplished by using a non-equilibrium AC Josephson measurement
 \cite{Sengupta-2001}. In principle,  protecting the fermion parity by a charging energy would allow the observation 
of the fractional Josephson effect in equilibrium. Observation of the fractional Josephson effect protected by $E_C$ 
y cannot occur in previously proposed linear Josephson junctions\cite{Kitaev-1D,Sengupta-2001,platero, meyer, nazarov,aguado}, 
which always have an additional pair of uncoupled MFs contributing to the fermion parity.  The loop geometry in 
Fig.~\ref{Fig2} would in principle allow the $4\pi$- periodic current phase relationship to be measured. However, such a current phase 
relation, would be relatively difficult to measure since the Josephson current would have to be measured in a closed loop circuit.
Finally, we note that the $4\pi$-periodicity in both the non-local transport and the Josephson case does not violate the Byers-Yang 
theorem \cite{byersyang} because of the long ranged Coulomb charging energy $E_C$, which is not accounted for 
in the BCS mean-field theory.

\paragraph{Summary and Conclusion:} In this paper we have proposed a scheme for uniquely identifying the Majorana assisted non-local electron
 tunneling between two MFs at the ends of a wire in the TS
phase. In principle, such a non-local transfer of electrons may be observable by a coincidence measurement
 \cite{sumanta,semenoff,gong}.
However, as we have shown here that the Majorana assisted electron tunneling process using  either a coincidence detection \cite{sumanta,gong} or 
by measuring the transconductance with a charging energy \cite{fu}, while interesting, cannot be taken as a definitive signature of MF modes
 because even conventional near-zero energy states trapped near the spatially separated leads can also produce such non-local signature. Instead 
we have proposed an interferometry experiment \cite{fu} appropriately generalized to geometries without edge modes.
We have shown that such a measurement can distinguish conventional and Majorana zero modes. 
 Our proposed non-local correlation experiment in terms of tunneling, which requires the inclusion of charging energy to fix the fermion parity,
 provides a direct verification of
the non-locality  of MFs in TS wires. We emphasize that the non-locality of the end state MFs arises from the non-locality of the 
fermion parity, which is unique to topological systems and cannot be emulated by conventional systems \cite{berg}.

J. D. S acknowledges support from the Harvard Quantum
Optics Center. S. T. would like to thank DARPAMTO,
Grant No. FA9550-10-1-0497 and NSF, Grant No.
PHY-1104527 for support. B. G. S. is supported by a Simons Fellowship through Harvard University.  We acknowledge valuable conversations with
Bertrand Halperin, Liang Fu and Charlie Marcus.

\appendix
\section{Appendix Sec. I: Absence of non-local correlations in the non-topological case}
We mention in the main text that the non-local correlation $G_R(\tau;ab)$ must vanish in the non-topological case. A 
particularly counter-intuitive case is
 the non-topological situation with two MFs at each end of a wire (as in Ref.~\onlinecite{beri}) where the wires 
are not coupled to each other. Such a system can in
 principle be prepared in a non-local initial state so that our current proposal gives a positive result by preparing 
each wire in a definite fermion parity state. However, as shown in the rest of the section,
 decoherence induced by coupling to the leads disfavors this non-local initial state and washes out any signature of non-locality in the 
present experiment.  Hence, although a weakly coupled pair of MFs (or a conventional zero-energy state) at each end may still show a ZBCP at 
or near zero bias at finite temperature,
 our interference measurement will correctly demonstrate that the system is smoothly connected to a non-topological phase.

 In the present section of the appendix, we show
that coupling to a Fermionic bath leads to such desctruction of non-local correlations. We
accomplish this in two sub-sections. In the first subsection, we show that when placed in contact
with a fermion bath in thermal equilibrium, the system reaches a
unique thermal equilibrium state. In the second subsection, we show that the system in thermal equilibrium
cannot have any non-local correlations of the type that might lead to the measured interference
or non-local tunneling amplitude.

\subsection{ Equilibration in the non-interacting case}
Let us now discuss equilibration of the ground state in the simple case of non-interacting fermions.
The non-interacting fermion problem can be solved in terms of
\begin{align}
&\rho_{ab}(t)=i\expect{\gamma_a(t)\gamma_b(t)}.
\end{align}
The Hamiltonian in terms of Majorana operators is written as
\begin{equation}
H=i\sum_{a,b}h_{a,b}\gamma_a \gamma_b.
\end{equation}
The equation of motion is written as
\begin{equation}
\partial_t \gamma_a = h_{a,b}\gamma_b,
\end{equation}
with $h_{a,b}$ being anti-symmetric. The time-evolution of $\gamma_a(t)$ is written as
\begin{equation}
\gamma(t)=G^{(R)}(t)\gamma(0),
\end{equation}
where $G^{(R)}(t)=e^{h t}$ for $t>0$. Expanding $G^{(R)}(t)$ in eigenstates
\begin{align}
&G^{(R)}(t)=\sum_n \psi_n \psi_n^\dagger e^{i\varepsilon_n t}\nonumber\\
&=\sum_n \psi_n \psi_n^\dagger \int_{-\infty}^{\infty}\frac{e^{i\omega t}}{\omega-\varepsilon_n-i\delta}=\int_{-\infty}^{\infty}\frac{e^{i\omega t}}{\omega-i h-i\delta}.
\end{align}
The time evolution of the density matrix is written as
\begin{equation}
\rho(t)=G^{(R)}(t)\rho(0) G^{(A)}(t).
\end{equation}

Assuming that we start with a density matrix that is diagonal between system $(s)$ and bath $(b)$ (i.e. lead),
the system density matrix
\begin{equation}
\rho_s(t)=G^{(R)}_{s}(t)\rho_s(0)G^{(A)}_{s}(t)+G^{(R)}_{sb}(t)\rho_b(0)G^{(A)}_{bs}(t).\label{eq:rhos}
\end{equation}

Assuming the system-bath structure for the Hamiltonian and defining $g_b^{(R,A)}=(\omega\mp i\delta-H_{b})^{-1}$, one can write
the final system Green function as
\begin{align}
&G^{(R,A)}_s(\omega)=(\omega\mp i\delta -H_{sb}g^{(R,A)}_b(\omega)H_{bs})^{-1}.
\end{align}
The system bath interaction Green function can be written as
\begin{align}
&G^{(R,A)}_{sb}(\omega)=G_s^{(R,A)}(\omega)H_{sb}g^{(R,A)}_b(\omega)\nonumber\\
&G^{(R,A)}_{bs}(\omega)=g^{(R,A)}_b(\omega)H_{bs}G_s^{(R,A)}(\omega).\label{eq:Gsb}
\end{align}
Provided the anti-hermitean part of the fermion-self-energy
\begin{equation}
\Sigma^{(R,A)}_s(\omega)=Im[H_{sb}g^{(R,A)}_b(\omega)H_{bs}]
\end{equation}
does not have a null-space (i.e. a space of zero-eigenvalues) the Green function $G^{(R,A)}_s(t)$ can be assumed
to decay exponentially in time. Therefore it follows from Eq.~\ref{eq:rhos} that the first term must vanish.
To evaluate the second term in Eq.~\ref{eq:rhos}, we note that from Eq.~\ref{eq:Gsb}, it is clear that
\begin{equation}
G^{(R,A)}_{sb}(t)=\int d\tau G_s(t-\tau)H_{sb}g_b(\tau).
\end{equation}
Since $G_s(t-\tau)$ is exponentially decaying, the $\Theta(\tau)$ component of $g_b$ is not relevant and one can approximate
\begin{align}
&G^{(R,A)}_{sb}(t)\approx \int d\tau G_s(t-\tau)H_{sb}e^{-h_b\tau}\nonumber\\
&=\int_{0}^{\infty} d\tau G_s(\tau)H_{sb}e^{-h_b(t-\tau)}
\end{align}
up to exponential factors. Applying this identity to Eq.~\ref{eq:rhos} we obtain
\begin{align}
&\rho_s(t)\nonumber\\
&\approx \int_0^\infty d\tau_1 d\tau_2 G_s^{(R)}(\tau_1)H_{sb}e^{-h_b(t-\tau_1)}\rho_b(0)e^{h_b(t-\tau_2)}H_{bs}G_s^{(A)}(\tau_2)\nonumber\\
&=\int_{-\infty}^\infty d\tau_1 d\tau_2 G_s^{(R)}(\tau_1)H_{sb}\rho_b(0)e^{h_b(\tau_1-\tau_2)}H_{bs}G_s^{(A)}(\tau_2),
\end{align}
which is a $t$-independent asymptotic value.
Transforming to frequency space,
\begin{align}
&\rho_s(t)\approx \int_{-\infty}^\infty d\omega f(\omega)G_s^{(R)}(\omega)H_{sb}A_b(\omega)H_{bs}G_s^{(A)}(\omega).
\end{align}
Considering the integrand at $\omega$ away from a pole of $g_s(\omega)$,
\begin{align}
&G_s^{(R)}(\omega)H_{sb}A_b(\omega)H_{bs}G_s^{(A)}(\omega)\nonumber\\
&=i G_s^{(R)}(\omega)H_{sb}\{g^{(R)}_b(\omega)-g^{(A)}_b(\omega)\}H_{bs}G_s^{(A)}(\omega)\nonumber\\
&=i G_s^{(R)}(\omega)\{\Sigma^{(R)}_s(\omega)-\Sigma^{(A)}_s(\omega)\}G_s^{(A)}(\omega)\nonumber\\
&=i \{G_s^{(R)}(\omega)-G_s^{(A)}(\omega)\},
\end{align}
which can be checked using Dyson's equation.
Therefore, in the absence of a protected null-space in the dissipative part of the self-energy $\Sigma_s(\omega)$,
the density matrix equilibrates to the grand-canonical thermal equilibrium value
\begin{align}
&\rho_s(t)\approx i\int_{-\infty}^\infty d\omega f(\omega)\{G_s^{(R)}(\omega)-G_s^{(A)}(\omega)\}.
\end{align}

\subsection{ Correlations in the grand-canonical thermal state}
Let us now discuss correlations in the grand-canonical thermal state.
First we prove a general lemma: Consider the grand-canonical thermal partition function of a system of Majorana fermions
(or fermions) which can be partitioned into two parts $L$ and $R$, so that the Hamiltonian has a symmetry $\gamma_{L,a}\rightarrow -\gamma_{L,a}$ for all MFs $\gamma_{L,a}$ on the left half of the system $L$. Then all correlation functions $\expect{\gamma_{L,a}\gamma_{R,b}}=0$.  To prove this simply apply the transformation $\gamma_{L,a}\rightarrow -U\gamma_{L,a}U^\dagger$ to
\begin{align}
&\expect{\gamma_{L,a}\gamma_{R,b}}\propto Tr[\gamma_{L,a}\gamma_{R,b}e^{-\beta H}]\nonumber\\
&=Tr[U\gamma_{L,a}U^\dagger\gamma_{R,b} Ue^{-\beta H}U^\dagger]\nonumber\\
&=-Tr[\gamma_{L,a}\gamma_{R,b}e^{-\beta H}]=0.
\end{align}
Therefore all such non-local fermionic correlations must vanish in the thermal state.

As discussed before, in the topological state, the charging energy gives rise to a fermion parity dependent term
in the Hamiltonian
\begin{equation}
H_{nl}\propto i\gamma_a\gamma_b,
\end{equation}
which violates the conditions for the above lemma and leaves a non-zero value for this non-local correlator.
Therefore the presence of the Majorana fermion term is crucial for the appearance of a non-local correlation function
that can contribute to the interference. Such an interference will not happen even if the conventional end zero mode
is made of a pair of Majorana fermions. This is because, as discussed in the previous sub-section, coupling Majorana
fermions to leads causes them to thermalize into the grand-canonical ensemble. Once they have thermalized in this ensemble,
$H_{nl}$ violates the conditions of the lemma and generates a non-local fermion correlation. The long-range Coulomb interaction
for conventional fermionic modes does not lead to the violation of the conditions of the lemma and does not
lead to any long range correlation. Following the rest of the argument in the text, it is clear that once non-local Green 
functions are absent, there can be no non-local correlation.

\section{Appendix Sec II: Details of nanowire and loop model}
Let us discuss the specific models for the nanowire and the loop. We use the 
standard Hamiltonian \cite{Long-PRB} for a spin-orbit coupled nanowire which can support 
both topological and non-topological phases.
The Hamiltonian is written as 
\begin{align}
&H_{NW}=\sum_n \ket{n}\{(2t-\mu)\tau_z+V_z\sigma_z+\Delta\tau_x\}\bra{n}\nonumber\\
&+[\ket{n}[-t+i\alpha\sigma_y]\bra{n+1}+h.c],
\end{align}
where $\sigma_{x,z}$ and $\tau_{x,z}$ are Pauli matrices representing the spin and particle-hole 
degrees of freedom. The state $\ket{n}$ represents quasiparticles on site $n$ with the 
spin and particle-hole index suppressed. 
Choosing parameters $t=$, $\alpha=$, $V_Z=$, $\Delta=$ corresponding to the relevant 
effective mass, spin-orbit and g-factor for InSb\cite{} we can obtain zero energy Majorana 
modes in the conventional phase .

The model for the loop self-energy  tunneling matrix $\Gamma$ associated with the loop, we need to compute the 
matrix elements $\rho_{ij}$ of the density matrix of the loop. To do this, we consider a simple 
tight-binding Hamiltonian 
for the loop, which does not include Zeeman or spin-orbit and is written as 
\begin{align}
H_{loop}=\sum_n \ket{n}\{(2t-\mu)\tau_z+i\zeta_n\}\bra{n}-t[\ket{n}\bra{n+1}+h.c],
\end{align}
where $\zeta_n$ characterizes the imaginary self-energy from the connection of the normal lead 
to the loop in the middle of the normal lead segment.
The density matrix is computed numerically as 
\begin{align}
&\rho_{ij}=Im[\expect{i|(H_{loop}+i\delta)|j}].
\end{align}
The parameters $\Gamma$ for the conductance is calculation is $\Gamma_0=t_a^2\rho_{aa}$ 
and $\Gamma_1=t_a t_b\rho_{ab}$. Numerically we find $\rho_{aa}=0.2$ and $\rho_{ab}=0.1$.

\section{Appendix Sec. III: Conductance calculation}
The main ingredient in the conductance equation in Eq.~[9] of the main-text is the retarded Green function.
The Green function to be computed is the correlation function 
\begin{align}
&G^{R}_{mn}(\tau)=-i\Theta(\tau)\frac{\expect{[c^\dagger_m(\tau),c_n(0)] e^{-(\beta+i\tau) F_{P}\delta}}_0}{\expect{ e^{-\beta F_P\delta}}_0},
\end{align}  
with respect to $H_{BCS}$.

We note that we can analytically continue $\tau$ to imaginary time when computing the 
conductance in Eq.~$[9]$ of the main text.
Let us expand the exponentials in the Green function in Eq.~$[5]$ in the main text
\begin{align}
&G^{R}_{mn}(i\tau)=-i\Theta(i\tau)\frac{\expect{[c^\dagger_m(\tau),c_n(0)] e^{-(\beta+\tau) F_{P}\delta}}_0}{\expect{ e^{-\beta F_P\delta}}_0},
\end{align}  
using the relation $e^{i a F_P}=\cos{a}+F_P\sin{a}$ so that 
\begin{align}
&\expect{[c^\dagger_m(\tau),c_n(0)] e^{-(\beta+\tau) F_{P}\delta}}_0\nonumber\\
&=\expect{[c^\dagger_m(\tau),c_n(0)] }_0\cosh{(\tau-\beta)\delta}\nonumber\\
&+\expect{[c^\dagger_m(\tau),c_n(0)] F_{P}}_0\sinh{(\tau-\beta)\delta}
\end{align}
and 
\begin{align}
&\expect{ e^{-\beta F_P\delta}}_0=\cosh{\beta\delta}+\expect{F_P}_0\sinh{\beta\delta}.
\end{align}
We will consider the above Green function in the long $\tau$ limit, so that the fermion parity $F_P$ is a product of the 
fermion parity of the end modes (both topological and non-topological zero modes). Splitting up the end modes (whether 
zero energy or not) into Majorana modes $\gamma_{j,L}$ at the left end and $\gamma_{j,R}$ at the right end we can write the 
fermion parity $F_P\propto\prod_j \gamma_{j,L}\gamma_{j,R}$. Furthermore at long imaginary times $\tau$ i.e. low energies, 
we expect the fermion correlator to factorize between the left and the right end (i.e. no correlations from the non-interacting BCS 
Hamiltonian).  
As a result, for $m,n$ at different ends, in the non-topological case, the numerator of $G$ vanishes at long imaginary times 
since there will be an odd number of fermion operators at each end.

In the topological case with a single Majorana at each end the fermion parity is written as $F_P=i\gamma_a\gamma_b$. Computing 
the Green function $G_{ab}$, we note that using the factorization of the correlator at large $\tau$, we get $\expect{[c^\dagger_m(\tau),c_n(0)] }_0=0$ 
and 
\begin{align}
&\expect{[c^\dagger_a(\tau),c_b(0)] F_{P}}_0=2 u_a u_b^*,
\end{align}
where we've used the fact that the MMs are at zero energy and $u_{a,b}$ are coefficients of Bogoliubov operators 
defined by $\gamma_{a,b}=u_{a,b}c_{a,b}^\dagger+\dots$. Substituting the Green function in the topological  case is written as 
\begin{align}
&G^{R}_{ab}(i\tau)=-i2 u_a u_b^*\Theta(i\tau)\frac{\sinh{(\tau-\beta)\delta}}{\cosh{\beta\delta}}.
\end{align}  
Analytically continuing back, we get the Green function in real time (and therefore frequency)
\begin{align}
&G^{R}_{ab}(\tau)=-i2 u_a u_b^*\Theta(\tau)\frac{\sinh{(i\tau-\beta)\delta}}{\cosh{\beta\delta}}.
\end{align}  
Fourier transforming and taking the imaginary part leads to the relation 
\begin{align}
&\rho_{ab}(\omega)=\frac{|u_a u_b|}{\cosh{\beta\delta}}[e^{\beta\delta}\delta(\omega-\delta)-e^{-\beta\delta}\delta(\omega+\delta)].
\end{align}
The $\varphi=\pi\Phi/\Phi_0$-dependent contribution to the conductance in Eq.~[9] in the main text is given by 
\begin{align}
&\sigma_{ab}=\frac{|u_a u_b|}{\cosh{\beta\delta}}\int d\omega \textrm{sech}^2{\frac{\beta\omega}{2}}[e^{\beta\delta}\delta(\omega-\delta)-e^{-\beta\delta}\delta(\omega+\delta)]\\
&=|u_a u_b|\textrm{tanh}{\beta\delta} \textrm{sech}^2{\frac{\beta\delta}{2}}.
\end{align}
Considering the conductance at the same end i.e. $\sigma_{aa}$, which is the $\varphi=\pi\Phi/\Phi_0$-independent 
term in Eq.~[9] in the main text, we note that 
\begin{align}
&G^{R}_{aa}(\tau)=-i2 |u_a|^2 \Theta(\tau)\frac{\cosh{(i\tau-\beta)\delta}}{\cosh{\beta\delta}}.
\end{align}  
Correspondingly the spectrum is  
\begin{align}
&\rho_{aa}(\omega)=\frac{|u_a|^2}{\cosh{\beta\delta}}[e^{\beta\delta}\delta(\omega-\delta)+e^{-\beta\delta}\delta(\omega+\delta)].
\end{align}
The corresponding conductance is given by 
\begin{align}
&\sigma_{aa}=T|u_a|^2 \textrm{sech}^2{\frac{\beta\delta}{2}}.
\end{align}
We will compute the Green function in the limit that impurities are chosen so that all the end sub-gap 
modes are at zero-energy. 

Assuming $|u_a|=|u_b|=|u|$ the diagonal conductance can be written as 
$\sigma_0\equiv \sigma_{aa}=\Gamma_0 |u|^2$, where $\Gamma_0$ is the coupling self-energy 
to the leads defined in the main text. We use $\sigma_0$ as the scale of the zero bias conductance 
measured at the end of the wire.
 The non-local conductance is obtained as  $\sigma_{ab}=\sigma_0 \textrm{tanh}{\beta\delta}$ 
so that substituting into Eq.~[9] in the main text, the total conductance is written as 
\begin{align}
&\sigma(\varphi)=\sigma_0[1+\lambda\tanh{\beta\delta/2}\cos{\varphi/2}].
\end{align}


\begin{thebibliography}
\bibitem{Supplementary2}Sec. II of the Appendix gives details of the nanowire and loop model, which are standard.
\end{thebibliography}
\end{document}